\documentclass{article}
\usepackage[T1]{fontenc}
\usepackage{charter}
\usepackage[utf8]{inputenc}
\usepackage[a4paper,bindingoffset=0in,%
            left=1in,right=1in,top=0.5in,bottom=0.5in,%
            footskip=.25in]{geometry}

\title{\textbf{A Systematic Paradigm for Detecting, Surfacing, and Characterizing Heterogeneous Treatment Effects (HTE)}}
\author{John Cai, Weinan Wang @ Snap Inc.}
\date{\vspace{-5ex}}
\usepackage{natbib}
\usepackage{graphicx}
\usepackage{amsmath}
\usepackage{bbm}
\usepackage{amssymb}
\usepackage[dvipsnames]{xcolor}
\usepackage{amsthm}
\usepackage{caption} 
\usepackage{parnotes}
\usepackage{lmodern}
\usepackage{amsmath}
\usepackage{physics}
\usepackage{scalerel,stackengine}
\stackMath
\newcommand\reallywidehat[1]{%
\savestack{\tmpbox}{\stretchto{%
  \scaleto{%
    \scalerel*[\widthof{\ensuremath{#1}}]{\kern-.6pt\bigwedge\kern-.6pt}%
    {\rule[-\textheight/2]{1ex}{\textheight}}
  }{\textheight}%
}{0.5ex}}%
\stackon[1pt]{#1}{\tmpbox}%
}

\usepackage{scrextend}
\changefontsizes[10pt]{8pt}
\usepackage[toc,page]{appendix}

\theoremstyle{plain} 
\newtheorem{theorem}{Theorem}

\theoremstyle{definition}

\theoremstyle{remark}

\begin{document}

\maketitle

\section{Motivation and Introduction}

At Snap, we run thousands of online controlled experiments to evaluate how feature changes affect a diverse range of users. Experimenters often propose a set of metrics (or one) as the \textbf{OEC}\parnote{overall evaluation criteria} for success measurement and examine the \textbf{ATE}\parnote{average treatment effect} or \textbf{QTE}\parnote{quantile treatment effect} at pre-specified locations. In addition, to gauge how different users are affected by a treatment, we often use a pre-defined set of breakdowns (e.g., countries, age bucket, gender, platform, etc.) to estimate the \textbf{HTE}\parnote{heterogeneous treatment effect}.

However, this manual approach is not only cumbersome but also subject to the experimenter's cognitive biases. Without a principled approach for global detection, experimenters waste time chasing down HTE when it does not exist. Without a way to surface and rank all HTE dimensions, experimenters rely on their pre-conceived notions of which dimensions are important, a practice that leads to confirmation bias and missed discoveries. Moreover, we also do not have a good way to precisely quantify how much each breakdown contributes to HTE. Last but not least, we need a systematic understanding of the distribution of such treatment effects beyond the average and certain pre-specified quantiles.

In this paper, we aim to tackle the aforementioned problems to enable a deeper understanding of HTE:
\begin{itemize}
\item
\textbf{Detection:} Global test to determine whether there’s HTE present or not.
\item
\textbf{Surfacing:} Systematically decompose treatment effect variation, measure how much variation can be explained by known breakdowns, and surface the HTE dimensions that explain most of the variation.
\item
\textbf{Characterization:} Model the distribution of individual treatment effects as a mixture of different conditional distributions, enabling further interpretable understanding of user level treatment effects.
\end{itemize}

\section{Related Work}

A popular model-based approach to detect HTE is to test if the conditional average treatment effects (\textbf{CATE}) are statistically different from each other when defined over a set of covariates (\cite{chang2015nonparametric}). However, this approach tests for whether HTE between subgroups exists rather than whether \textit{any} HTE exists. Treatment effect variation (\textbf{TEV}) within a subgroup defined by a covariate will not cause the null to be rejected (\cite{crump2008nonparametric}). 

To have an omnibus test on the existence of TEV in general, we study the model-free approaches proposed by \cite{Ding16}. In particular, they propose using Fisher's Randomization Test (FRT) (\cite{rosenbaum2002covariance}), with modifications to make the test robust to estimation error of the ATE by taking the maximum p-value over a specified range. At the same time, \cite{Ding16} propose using a Kurtosis test that appeals to higher-order moments to construct an asymptotically valid test for the difference in log-variances of treatment and control. We find these model-free approaches more applicable as a general detection tool prior to the specification of observed covariates to define subgroups and explain HTE.

For HTE surfacing, we use the $R^2$-like measure proposed by \cite{ding2019decomposing}. They construct the $R^2$-like measure by estimating explained TEV, and forming an upper bound on the idiosyncratic TEV. As pointed out by \cite{fan2010sharp}, we can only observe the marginal distributions of treatment outcomes and control outcomes for a disjoint set of individuals, but not the dependence structure between treatment and control outcomes. Hence, following \cite{heckman1997making}, in the absence of information on the dependence structure, we can use Fr\'echet-Hoeffding bounds on the joint distribution of treatment and control outcomes to form sharp bounds on the treatment effect distribution. The distributional bounds can in turn be used to bound the idiosyncratic TEV and the $R^2$-like measure as shown in \cite{ding2019decomposing}.

\section{Methodology and Examples at Snap}
In this section, we briefly describe the methodologies involved in solving the three aforementioned problems. Without loss of generality, we focus on the case of a randomized controlled experiment with treatment indicator $Z_i$ and continuous outcome $Y_i$ for the $i$-th user. Let $N$ be the total number of users in the study, with $N_1$ in treatment and $N-N_1=N_0$ in control. Invoking the SUTVA condition, each individual's observed outcome can be denoted as: $Y_{i}^{\text{obs}}=Z_iY_i(1)+(1-Z_i)Y_i(0)$. Define the \textbf{ITE}\parnote{individual treatment effect} the usual way as $\tau_i=Y_i(1)-Y_i(0)$.

\subsection{Global HTE Detection}
\subsubsection{Methodology}
In general, following \cite{Cox84}, we can assess treatment effect heterogeneity by examining the marginal variances of the treatment and control groups' respective outcomes. Specifically, if the overall treatment effect is constant, i.e. $Y_i(1)=Y_i(0)+\tau$, then $\text{Var}(Y_i(1))=\text{Var}(Y_i(0))$. This makes the test statistic of variance ratio $t_{\text{var}}=\hat{\sigma}_1^2/\hat{\sigma}_0^2$ sensible.

However, we would not expect the $t_\text{var}$ to follow an $F$-distribution since the marginal distributions of the potential outcomes are not normal for metrics at Snap. Hence, we appeal to a more general theorem that uses higher order moments:

\begin{theorem}
Assume that $Y(z)$ has finite kurtosis $\kappa_z$.  Under the null of equal variance,
$$
t_{\text{kvar}}=\frac{\log\hat{\sigma}_1^2-\log{\hat{\sigma}_0^2}}{\sqrt{(\hat{\kappa}_1-1)/N_1+(\hat{\kappa}_0-1)/N_0}}\xrightarrow{d}\mathcal{N}(0,1)
$$
\end{theorem}

Additionally, we explore the use of a bootstrapped F-test for the variance of the mean. While the metrics are unlikely to be normally distributed, we can appeal to the central limit theorem to compute bootstrapped means, which we can then compute the variance of. This variance would be tightly related to the variance of the metrics.

\subsubsection{Simulation Results for HTE Detection}
To demonstrate the power of this method, we run a small simulation comparing the proposed method (\textbf{Kurtosis}) against the bootstrapped \textbf{F-test} (200 bootstraps) and the Fisher's Randomization coupled with KS Test (\textbf{FRT}) in \cite{Ding16}. Specifically, for a total of $N=5000$ users, we simulate AA period data for all as negative binomial distribution with parameter $A_i\sim \mathcal{NB}(\mu=200,\phi=0.15)$, and AB period data for control as $Y_i(0)\sim A_i+\mathcal{N}(0,1)$, and treatment effect $\tau_i\sim \mathcal{LN}(\mu=2,\sigma^2)$ and we vary $\sigma$ from 0 to 0.15, simulating scenarios where there's no treatment effect variation to strong treatment effect variation. A total of 500 simulations are conducted. We compare the power of these three procedures by utilizing the BH procedure with nominal FDR level at $\alpha=0.05$ on the resulting $p$-values and compare the proportion of detected significant results.
\begin{center}
\includegraphics[width=0.8\linewidth]{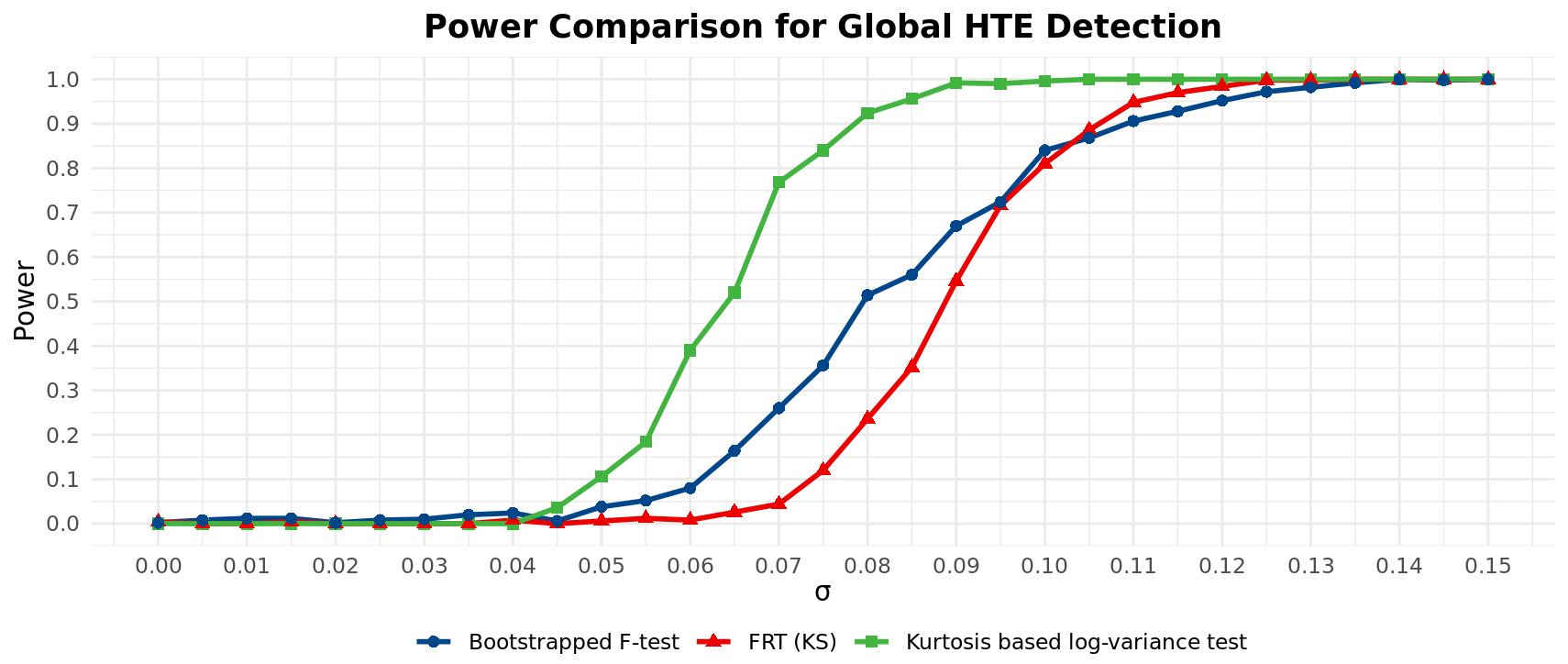}
\end{center}

Going forward, we use the kurtosis-based log variance test, as it has has better power and is also robust to non-normality. Formally, we compute the two-sided $p$-value: $2\Phi(-|t_{\text{kvar}}|)$ for global HTE detection on any given OEC.
\subsubsection{Empirical Results at Snap}
To check if the method is working as intended on real AB data, we apply the kurtosis-based log-variance test on AA and AB data. We anonymize all the metrics that we have computed the results on, and only make the distinction between whether a metric is a count-based metric (e.g. count of views) versus whether it is a days-based metric (e.g. number of active days). For the former, it is generally unbounded above 0, while for the latter, there is an upper bound in the number of days that have elapsed during the experiment.

\begin{center}
\includegraphics[width=1\linewidth]{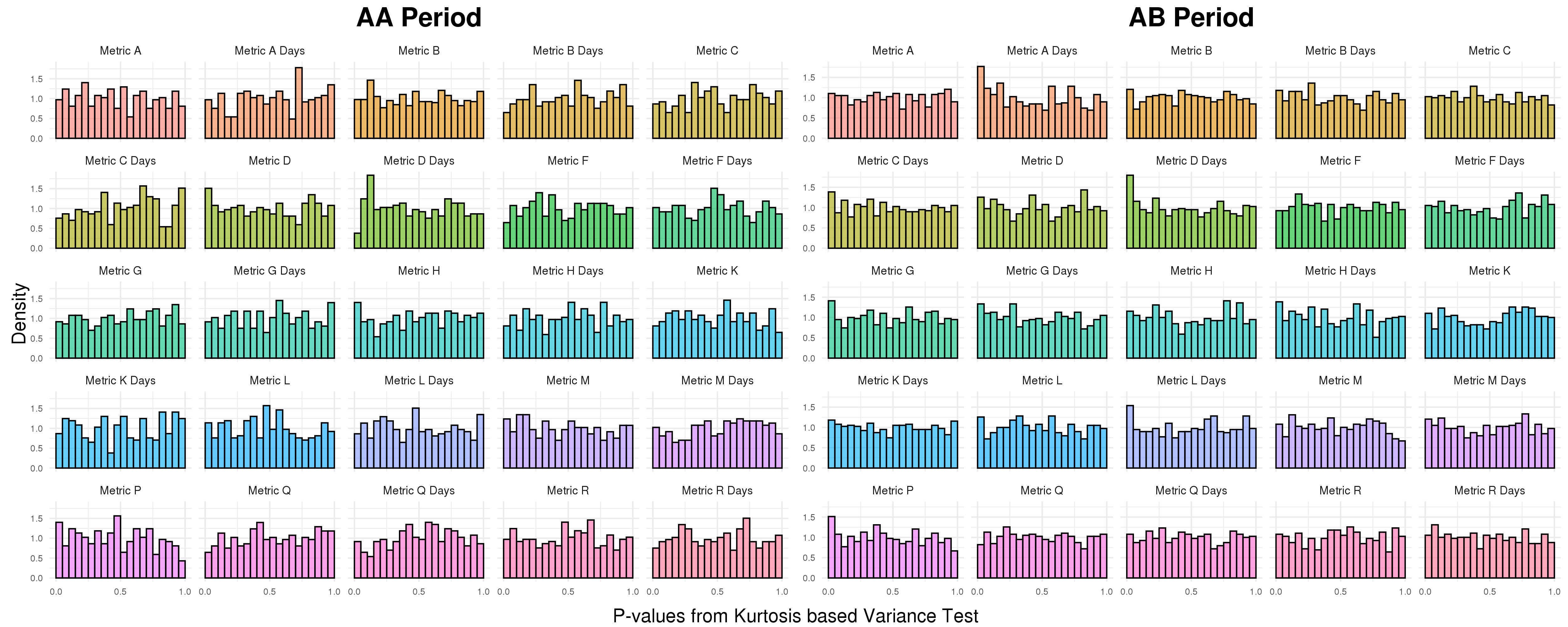}
\end{center}

From above, we observe that the AA p-values are generally uniformly distributed. In the AA period (pre-experiment), no treatment has been applied yet. Hence, there is no systematic difference in distribution between the treatment and control group. Therefore, we would expect the p-values to be uniformly distributed, as shown above.

For A/B data, we observe that days metric (e.g. Metric A Days, Metric D Days) tend to have a higher proportion of p-values $< 0.05$. Days metric measures the number of days that a user has been active over a given period. The heterogeneity in the metric could be attributed to its inherently bounded nature, as a user who is active for all the days of the experiment cannot increase their active days. 

To correct for false discoveries that occur due to the metrics' inherent distribution, it is important to apply false discovery control methods. We use the past distribution of the p-values for every metric, and \textbf{apply the Benjamini-Hochberg false discovery rate control procedure}  from \cite{benjamini1995controlling}. With this adjustment, we can accurately control the false discovery rates, and reduce the effects of a metric's inherent susceptibility to HTE.

\subsection{HTE Surfacing through Most Likely Breakdowns}
\subsubsection{Methodology }
To surface the dimensions that contribute the most towards treatment effect variation, we construct a $R^2$-like measure of the ratio of explained treatment effect variation to total treatment effect variation. This measure is similar in construction to the one described in \cite{ding2019decomposing}. For simplicity, suppose that the treatment effect was correlated with existing covariates $\boldsymbol{X_i^T}$, and that an individual takes on the following potential outcomes in control and treatment, i.e. $Y_i(0),\;Y_i(1)$; let $e_i$ and $\epsilon_i$ denote the error terms in the control outcomes and the treatment effect respectively.
$$
    \tau_i = \boldsymbol{X_i^T \beta} + \epsilon_i,\;
    Y_i(0) =   \boldsymbol{X_i^T \gamma} + e_i,\;
    Y_i(1) =  \boldsymbol{X_i^T \gamma}+ \tau_i + e_i =  \boldsymbol{X_i^T} (\boldsymbol{\gamma + \beta})  + \epsilon_i + e_i.
$$

We begin by estimating a model of how the control outcomes vary with $X_i$ using an OLS model. In our analysis, we focus on categorical variables. Hence, we estimate our model using the conditional means of each category.
$$
    \hat{Y_i}(0)= \boldsymbol {X_i^T \hat{\gamma}},\;
    \hat{Y_i}(1)= \boldsymbol{X_i^T}  ( \boldsymbol{\hat{\gamma} + \hat{\beta}}). 
$$

The \textbf{Total Treatment Effect Variation (TEV)} can be decomposed into two components, the explained and the idiosyncratic variation. To obtain the explained treatment effect variation, we directly estimate the variation in the conditional average treatment effects, which is the difference between the predicted treatment and control outcomes ($\hat{Y_i}(1) - \hat{Y_i}(0)$). 
\begin{align*}
     \text{Total TEV} &= \mathrm{Var} (\tau_i) =\mathrm{Var}( \boldsymbol {X_i^T \hat{\beta} }) + \mathrm{Var}(\epsilon_i), \;
    \text{Explained TEV} = \mathrm{Var}( \boldsymbol {X_i^T \hat{\beta} }) = \mathrm{Var}( \hat{Y_i}(1) - \hat{Y_i}(0)), \\
    \text{Idiosyncratic TEV} &=  \mathrm{Var}(\epsilon_i), \;
    R^2 = \frac{\text{Explained TEV}}{\text{Total TEV}} = \frac{\mathrm{Var}( \boldsymbol {X_i^T \hat{\beta} })}{\mathrm{Var}( \boldsymbol {X_i^T \hat{\beta} }) + \mathrm{Var}(\epsilon_i)}.
\end{align*}

Estimating the Idiosyncratic TEV $\mathrm{Var} (\epsilon_i)$ is more complicated, as we do not observe each individual under both treatment and outcome states. Hence, we use the bounds proposed in \cite{ding2019decomposing}, but with modifications to increase precision and rely on less stringent assumptions. Specifically, due to the nature of the randomized trials, we would expect balanced residual distributions between treatment and control on any strata. Instead of deriving the bounds of $R^2$ based on the rank matching of the global residual distributions, we match on each strata of $\boldsymbol{X_i^T}$. Furthermore, we use a direct matching process rather than using quantile approximations. The procedure is described below:

\begin{enumerate}
\item Obtain the residuals after removing the predicted outcomes from the actual outcomes. This gives us: $\hat{v}_i(0) = Y_i(0) - \boldsymbol {X_i^T \hat{\gamma}}$ and
$\hat{v}_i(1) = Y_i(1) - \boldsymbol{X_i^T}  ( \boldsymbol{\hat{\gamma} + \hat{\beta}}) $ . For treatment, it consists of both the treatment effect error term $\epsilon_i$ and the control outcome error term $e_i$, while for control, it consists of just the control outcome $e_i$.
\item Group all residuals using each category of $\boldsymbol{X_i^T}$. This leaves us with $M$ strata, where each strata has $N_{m_0}$ individuals in control and $N_{m_1}$ individuals in treatment.
\item Sort residuals within each strata s.t. $\hat{v}_k(0) \geq \hat{v}_{k-1}(0),\; \forall  k \in \{2,...,N_{m_0}\} $ and $\hat{v}_k(1) \geq \hat{v}_{k-1}(1),\; \forall k \in \{2,...,N_{m_1}\}$.

\item Map each residual to a rank $j$ s.t. $1 \leq j \leq N_{\text{min}} = \min(N_{m_0},N_{m_1})$ within each strata. WLOG\parnote{without loss of generality}, suppose that for each strata $N_{\text{min}} = N_{m_0}$. Then, define $S_j(1) = \{\hat{v}_k(1) \; | \, \left(\left \lfloor{k \cdot (N_{\text{min}} / (N_{m_1}+1))}\right \rfloor + 1 \right) = j \} $. The treatment group's residual used in counterfactual matching would then be the average of the set: $\tilde{v}_j = \frac{1}{|S_j|} \sum_{\hat{v}_k \in S_j}\hat{v}_k$.
\item Estimate the residual difference for both treatment and control: $\hat{d}_j(0) =  \tilde{v}_j(1) - \hat{v}_j(0)$, $\hat{d}_j(1) = \hat{v}_j(1) - \tilde{v}_j(0)$.    
\item Obtain $\mathrm{Var}(\hat{d}_i) = \frac{1}{N-1} \sum_{i=1}^{N} (\hat{d}_i )^2$, as $E(\hat{d}_i) = 0$ by construction. This is the discrete analogue of the Fr\'echet-Hoeffding bound when we only observe the marginal distributions: $ \frac{1}{N-1} \sum_{i=1}^{N} (\hat{d}_i )^2 \approx \int_0^1 \{F_1^{-1}(u) - F_0^{-1}(u) \}^2du \leq \mathrm{Var}(\epsilon_i)$ where $F_1^{-1}$ and $F_0^{-1} $ are the empirical CDF of $\hat{v}_j(1) $ and $\hat{v}_j(0) $ respectively (\cite{aronow2014sharp}).
\item Under local rank-preservation between $v_j(1) $ and $v_j(0)$ \textbf{within each strata}, we obtain the idiosyncratic TEV $\mathrm{Var(\epsilon_i)} = \mathrm{Var}(\hat{d}_i)$. Our assumption of local rank-preservation within each strata is less stringent than the assumption of global rank-preservation used in \cite{ding2019decomposing} (\textbf{Theorem 2,3}). As a corollary, our $R^2$ bounds are sharper.
\item Combine the idiosyncratic TEV with the explained TEV to obtain the upper bound of $R^2$.
\end{enumerate}

\subsubsection{Simulation Results for $R^2$ Computation}

To understand how well our method bounds the $R^2$ compared to the true $R^2$, we run a simulation where we know the counterfactual treatment and control outcomes. We also compare the performance of the stratified bound versus the unstratified bound (in \cite{ding2019decomposing}), and demonstrate that our bound is tighter.

To simulate our results, we assume that the control outcomes error $e_i$ are drawn from a negative binomial distribution of $NB(30,0.5)$. Negative binomial distributions are appropriate in modeling discrete count-based metrics (\cite{hilbe2011negative}). The control outcome is then modelled by: $Y_i(0) = \boldsymbol{X_i^T \gamma} + e_i$, with $X_i$ being a categorical variable with 4 possible values. In each category, we have 500,000 users. The treatment effect error term is then modeled using the marginal distribution $\epsilon_i  \sim NB(\mu_j, 0.5) - (\mu_j * 0.5) $ to center idiosyncratic treatment effects at 0. $\mu_j$ varies across different strata. 

Subsequently, we link the control outcome error $e_i$ and the treatment effect error $\epsilon_i$ using a Gaussian copula with varying levels of correlation $r$. Finally, we generate a treatment assignment vector $Z_i$. 
\begin{center}
\includegraphics[width=0.98\linewidth]{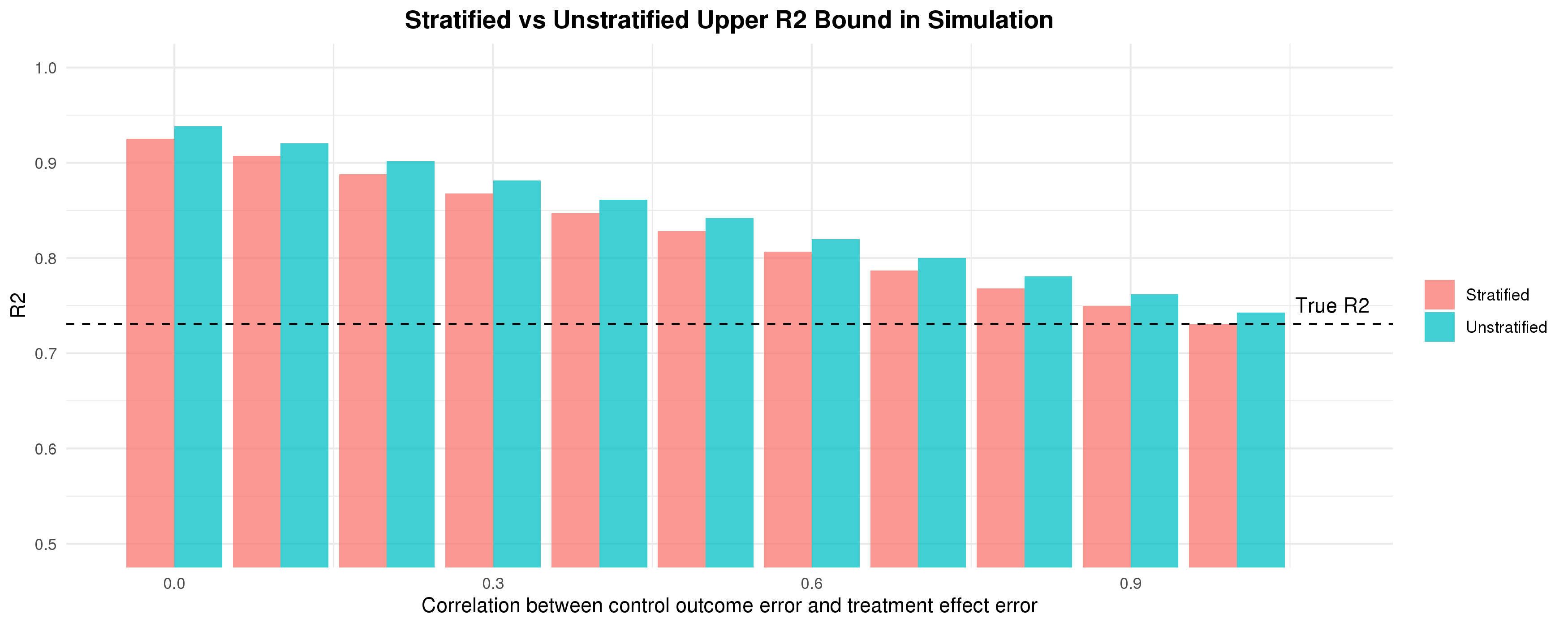}
\end{center}

From above, we have two key insights. First, we observe that the upper stratified bound is tighter than the unstratified bound across all ranges of the correlation. This is expected because we induce additional rank perturbations across the strata, as captured by the different treatment effect distributions simulated. Second, we observe that as the correlation of the error terms increases from 0 to 1, the bounds of the $R^2$ estimate converge to the True $R^2$. Intuitively, as the dependence captured by the Gaussian copula increases, the treatment effect will be more locally rank-preserving, which allows the locally stratified $R^2$ to converge to the true ground-truth $R^2$.

\subsubsection{HTE Surfacing applied at Snap }

Using an Ads ranking experiment conducted at Snap as illustration, we examine an anonymized metric that is shown to have global HTE through the kurtosis based variance test mentioned in the previous section, and rank the upper bound of the $R^2$ measured on known popular categorical breakdowns:
\begin{center}
\includegraphics[width=1.0\linewidth]{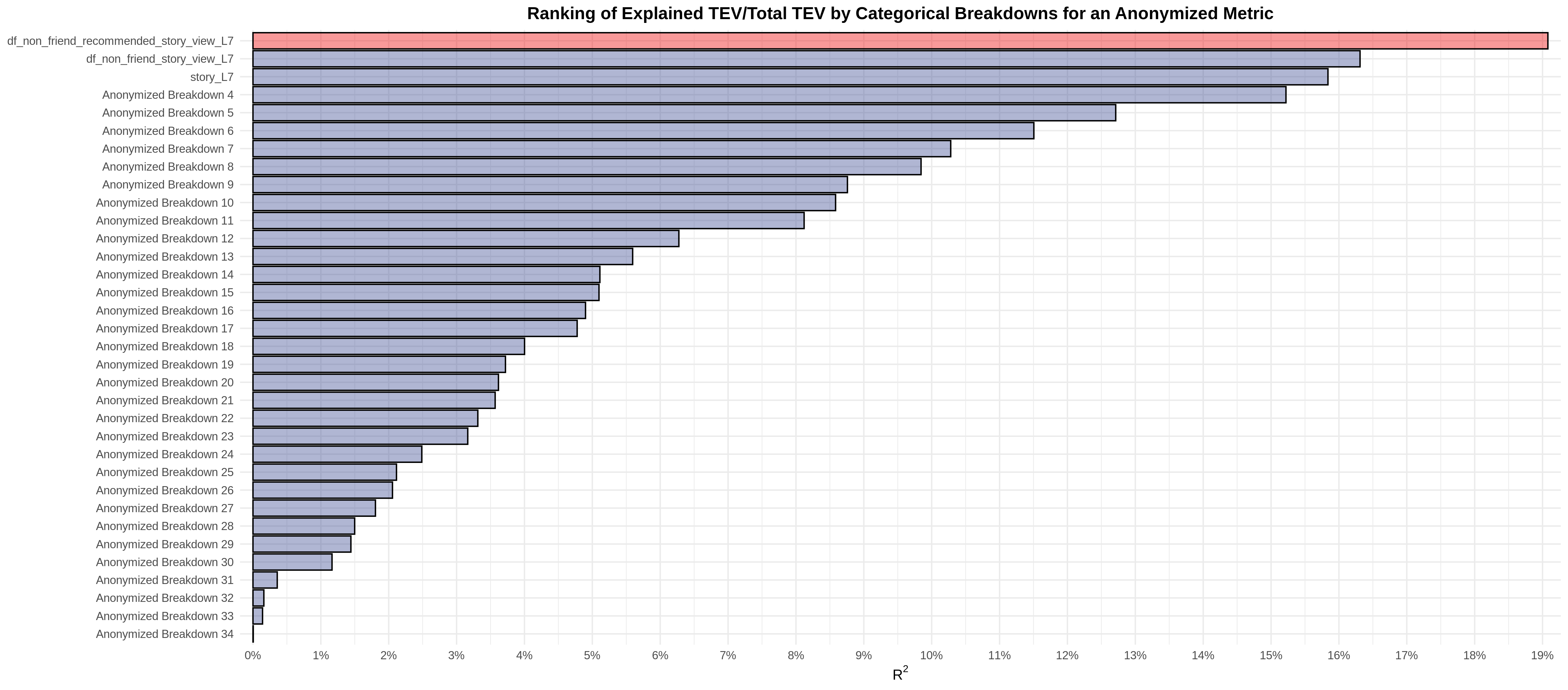}
\end{center}
From above, we see that  \textbf{df\_non\_friend\_recommended\_story\_view\_L7} ranks the highest in explanatory power for the treatment effect variation, with an upper bound  $R^2\approx 19\%$. This metric captures the number of days in the past week that a user views a recommended story. As the experiment was designed to better utilize signals from previously recommended content, past engagement with recommended content would naturally be highly correlated with the treatment effect. Other breakdowns that are tightly related to the top breakdown, like \textbf{df\_non\_friend\_story\_view\_L7} and \textbf{story\_L7} are also ranked second and third respectively. These are more generic versions of the top breakdown.

\subsection{Distributional Characterization of Individual Treatment Effects (ITE)}
Now, we have the essential tools for detecting the existence of HTE and finding the one-dimensional covariate that can explain the greatest amount of treatment effect variation. In this section, we go beyond the decomposition of the global ATE with a mixture of \textbf{CATE}\parnote{conditional average treatment effect}  on the selected covariate, and characterize the global ITE distribution as a mixture of conditional distributions for each strata.

As mentioned above, we already have a sensible estimate for the residual differences $\hat{d}_i(0),\hat{d}_i(1)$, for each user, we would approximate the ITE $\tau_i$ as a combination of the estimated CATE and the idiosyncratic treatment effect:
$$
\hat{\tau}_i=\pmb{X}_i^T\hat{\beta}+\hat{d}_i(0),\;\text{for }i \text{ in control};\; \hat{\tau}_i=\pmb{X}_i^T\hat{\beta}+\hat{d}_i(1),\;\text{for }i \text{ in treatment}.
$$

We can subsequently pool all estimates $\hat{\tau}_i$ together and surface conditional ITE distributions for each strata $\pmb{X}_i^T$, the vertical dashed lines indicate CATE (P5 to P95 are shown for visualization):
\begin{center}
\includegraphics[width=0.98\linewidth]{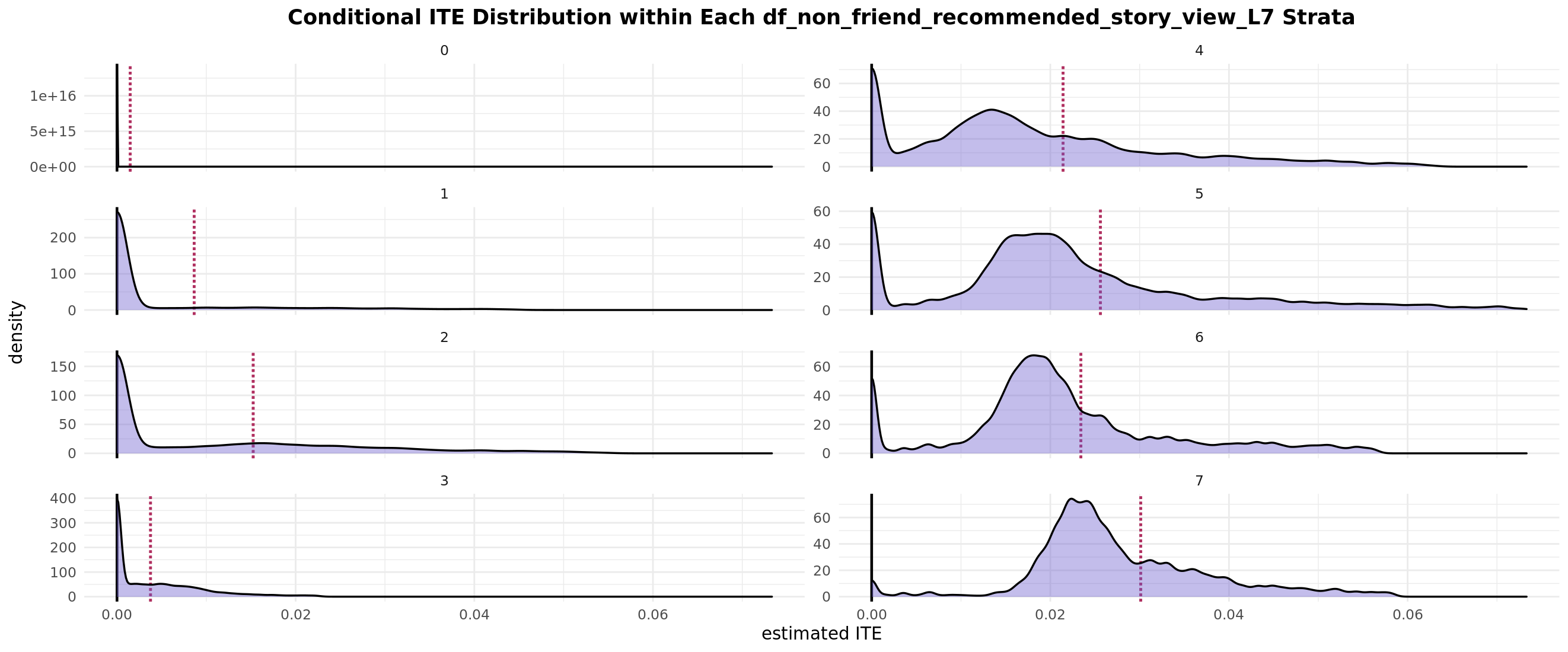}
\end{center}

As seen above, for the strata that df\_non\_friend\_recommended\_story\_view\_L7 is 0, the ITE distribution is largely a point mass near 0, indicating a very small treatment effect with small dispersion. This is contrasted with other conditional distributions, which have some dispersion above 0. Intuitively, users with no activity in the past do not have signals that the model in the treatment can use to recommend better content, resulting in negligible benefit to these users.

As we move upwards the stratification in previous engagement days from 1 to 7, we see the ITE distribution showing more of a mixture structure of positive impact and the 0 point mass, with the positive impact's mode increasing as well. We also observe a longer right-tail for distributions after previous engagement days reaches 4.

Furthermore, we can visualize the global ITE distribution as a weighted mixture of these empirical distributions within each strata:
\begin{center}
\includegraphics[width=0.98\linewidth]{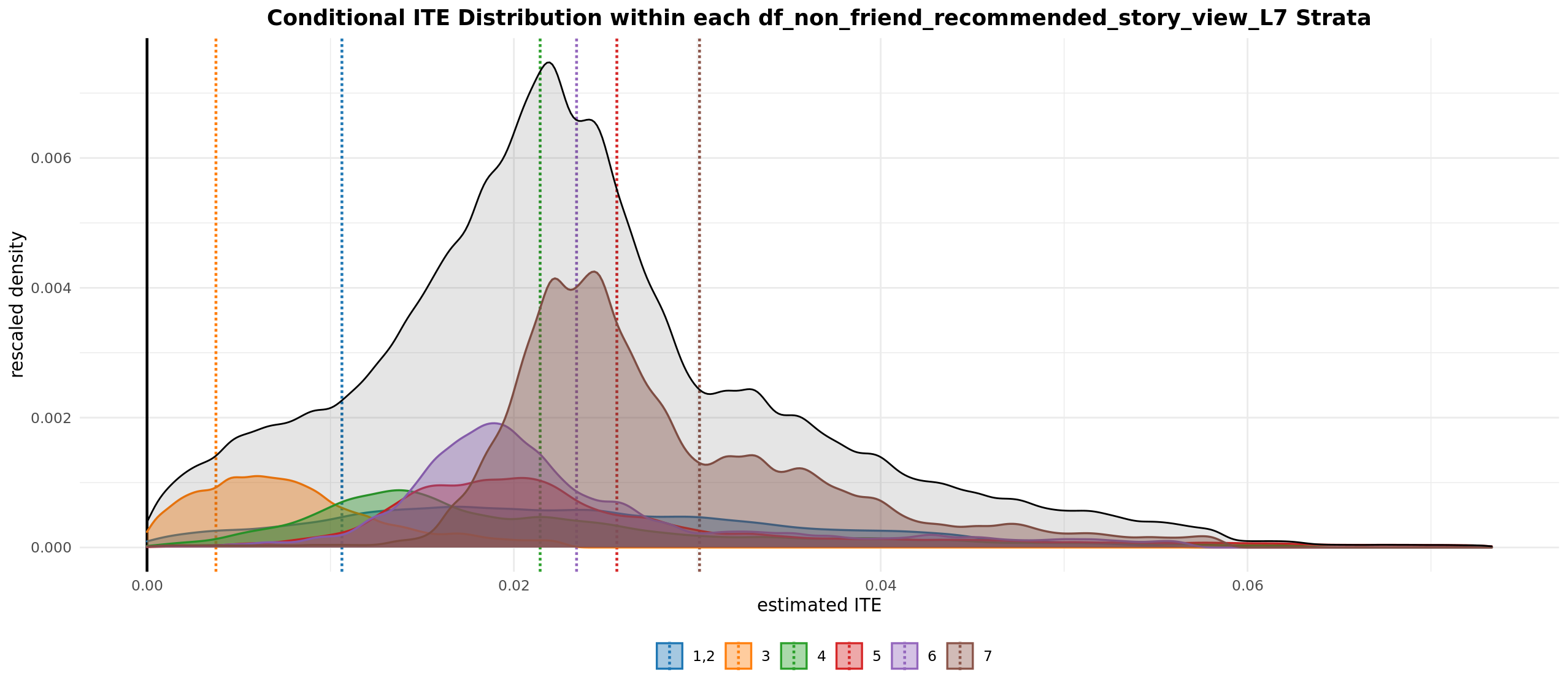}
\end{center}

For the sake of visualization, we removed the 0 cohort as it's mostly a point mass and thus less interesting, and group strata 1 and 2 together as they are very similar in shape. We clearly see the value in decomposing the distribution of treatment effect based on the strata, as the conditional distributions not only have wildly different means, but they also have highly varying modes and tail distributions. Interestingly, we observe that at the right tail, beyond the peak of the main distribution, most of the tail treatment effects come from strata 7. Users who benefit tremendously from the treatment are highly likely to be strata 7 users. Moreover, the results also suggest that further investigation to add sub-stratas on top of strata 5,6,7 could help to shed light into the composition of these conditional distributions.
\appendix
\section{Appendix}
\begin{theorem} \textbf{Global rank preservation implies local rank preservation}

 Let $x_i$ be a series of sorted values s.t. $x_i \geq x_{i-1} \; \forall \, i \in \{2,...,N \}$. Let $f(.)$ be a globally rank-preserving function s.t. $f(x_i) \geq f(x_{i-1}) \; \forall \, i \in \{2,...,N \}$.

Split the set of $Q = \{x_i \, | \, i \in \{2,...,N \} \}$ into any partition $P$, where $\cup_{j \in J} (P_j) = Q$ and $P_j \cap P_k = \emptyset$ for $j,k \in J$ with $j\neq k$. A locally-rank preserving function is a function $g(.)$ s.t. $ \forall \, x_n,x_m \in P$, $x_n \geq x_m \Rightarrow g(x_n) \geq g(x_m)$.

Since $f(.)$ preserves the rank globally, it can be easily seen that when comparing any $x_n, x_m$ within any partition $P$, whenever $x_n \geq x_m$, we will have $f(x_m) \geq f(x_m)$. Hence, global rank preservation implies local rank preservation.
\end{theorem}

\begin{theorem} \textbf{Local rank preservation does not imply global rank preservation}

Suppose that we have two partitions $P_j, P_k$ where $\inf(P_j) > \sup(P_k)$. Let our function $g(.)$ have a locally rank-preserving functional form below:
\[
g(x)= 
\begin{cases}
    x + \sup(P_j), & \text{if } x \in P_k\\
    x,              & \text{if } x \in P_j
\end{cases}
\]
Within each partition, we apply a monotonically increasing function, and is hence locally rank-preserving. However, when comparing across partitions, we can easily see that $g(x_j)< g(x_k)  \; \forall \, x_k \in P_k , x_j \in P_j $, even though $x_j \geq x_k $. This is in contradiction to the definition of global rank preservation, which requires that $x_j \geq x_k \Rightarrow g(x_j) \geq g(x_k) \; \forall \, x_j, x_k \in Q$.
\end{theorem}

\parnotes
\bibliographystyle{chicago}
\bibliography{references.bib}

\end{document}